# ETUDE DE L' OSCILLATEUR DE VAN der Pol GENERALISE PAR LA METHODE DU GROUPE DE RENORMALISATION

HINVI A. L., MONWANOU V. A. & CHABI OROU J. B.*

Institut de Mathématiques et de Sciences Physiques, BP : 613 Porto-Novo, Bénin.

Correspondance : Jean B. Chabi Orou jchabi@yahoo.fr

Résumé

La méthode du groupe de renormalisation est l'une des méthodes de perturbation singulière utilisée dans la recherche des comportements asymptotiques de solution des équations différentielles ordinaires. Dans ce papier, l'équation de l'oscillateur de VAN der Pol généralisé qui modélise beaucoup de phénomènes physiques est considérée. Un bref rappel de la technique est fait, ensuite elle est appliquée à l'équation de l'oscillateur de VAN der Pol généralisé pour faire ressortir sa solution asymptotique.

Mots clés : Oscillateur de VAN der Pol généralisé, groupe de renormalisation, solution asymptotique.

Abstract

The renormalization group method is one of the singular perturbation methods used in the research of the asymptotic behavior of solution of ordinary differential equations. In this paper, the equation of VAN der Pol generalized oscillator that models many physical phenomena is considered. A brief review of the technique is done and is applied to the generalized VAN der Pol oscillator to highlight its asymptotic solution.

1. **Introduction**

L'analyse du comportement asymptotique a joué un rôle important dans le domaine de la mathématique appliquée et de la physique théorique. Dans plusieurs cas les méthodes de perturbation régulière deviennent inapplicables en faveur des méthodes de perturbation singulière (CHEN L. Y. et al., 1994)-(HINCH E. J., 1991). Nous pouvons citer comme méthodes de perturbation singulière pour la résolution des équations différentielles ordinaires (EDO), les méthodes des échelles multiples, de la couche limite, des moyennes, WKB (BENDER C. M and ORSZAG S. A., 1978), (KEVORKIAN J. D. and CODE J. D., 1981), (NAYFEH A. H., 1973), (SMITH D. R., 1985) and (LOMOV S. A., 1992), la méthode de reconstitution (ROBERTS A. J., 1985), la théorie de variété centrée (CART J., 1981) and (CART J. and MUNCASTER R. M., 1983) etc. Chacune de ces méthodes possède ses propres inconvénients dans le mécanisme ou l'algorithme de son application. En effet, il est probablement, juste de dire que la pratique de l'analyse asymptotique est quelque chose de l'art.

La méthode du groupe de renormalisation qui fait l'objet de cette étude fut établi par CHEN L. Y. et al., (1994) pour les équations différentielles de la forme

$$\dot{x} = Fx + g(x, t, \epsilon); \ x \in \mathbb{R}^n, \qquad (1)$$

où $\epsilon$ est un paramètre positif infiniment petit.

Ils ont montré qu'elle unifie les méthodes de perturbation singulière énumérées ci-dessus (CHEN L. Y. et al., 1994). KUNIHIRO J. et al., (2000) a montré le lien qui existe entre la méthode des échelles multiples et la méthode du groupe de renormalisation.

L'analyse des (EDO) par la méthode du groupe de renormalisation est consacrée particulièrement à la construction des solutions uniformes ou globalement approximatives convergentes pour des variables indépendantes petites ou larges. Avec cette méthode, la renormalisation des constantes d'intégration permet de lever la divergence (suppressions des termes séculaires).

Cette technique de renormalisation fait apparaître une équation dite équation du groupe de renormalisation (EGR) mettant en jeu l'amplitude de l'oscillation. Cette dernière est plus facile à manipuler dans la détermination de l'amplitude qui rend stable le cycle limite.

La méthode du groupe de renormalisation est aussi applicable aux équations aux dérivées partielles tel que les équations de BARENBLATT G. I., (1979), les problèmes de propagation du front d'onde (les équations de réaction-diffusion) (PAQUETTE G., 1994) and (PAQUETTE G. et al., 1994).

CHIBA H. (2008b) aussi a montré le lien entre ces méthodes et a utilisé la méthode du groupe de renormalisation pour analyser le model de Kuramoto pour les oscillateurs couplés (CHEN L. Y. et al., 1994). Dans ce travail nous avons décidé d'étudier l'oscillateur de VAN der Pol dans sa forme générale par l'équation adimensionnée ci-dessous :

$$\ddot{x} + x - \epsilon(1 - ax^2 - b\dot{x}^2)\dot{x} = 0, \qquad (2)$$

avec $a, b, \epsilon$ des paramètres positifs réels et $\epsilon$ infiniment petit. Cette forme est dite générale parce qu'elle modélise un bon nombre d'oscillateurs grâce aux paramètres $a$ et $b$ sur lesquels on peut agir pour transformer cette équation aux formes particulières qu'on rencontre souvent dans la littérature et à d'autres catégories d'oscillateurs. Par exemple pour $a = 1$ et $b = 0$ on retrouve

$$\ddot{x} + x - \epsilon(1 - x^2)\dot{x} = 0, \qquad (3)$$

la forme la plus simple de l'équation de l'oscillateur de VAN der Pol rencontré dans littérature. On peut citer ACEVES A. et al., (1994) et PAQUETTE G. and OONO Y., (1994). De même pour $a = 0$ et $b = \frac{1}{3}$ on retrouve l'équation de l'oscillateur de Rayleigh

$$\ddot{x} + x - \epsilon\left(1 - \frac{1}{3}\dot{x}^2\right)\dot{x} = 0, \qquad (4)$$

étudiée par THOMAS C. B., (2005). De plus pour $a = 3$ et $b = 0$, on retrouve

$$\ddot{x} + x - \epsilon(1 - 3\dot{x}^2)\dot{x} = 0, \qquad (5)$$

qui est l'exemple $N°1$ traité par CHIBA H., (2008b). L'oscillateur de VAN der Pol généralisé est un oscillateur qui modélise beaucoup de phénomènes physiques comme en électronique, en mécanique, etc.

NAYFEH A. L., (1981) a fait ressortir les similarités, avantages, différences ainsi que les limites des méthodes de perturbation énumérés ci-dessus en les appliquant à des classes d'oscillateurs comme l'oscillateur de Rayleigh, l'équation de duffing, les oscillateurs linéaires, les systèmes avec non linéarités cubique et ou quadratique forcé ou non etc. Dans chaque cas d'oscillateur il compare les solutions analytiques trouvées par ces techniques à la solution exacte et en déduire la méthode la plus plausible. Il a étudié l'oscillateur de Rayleigh qui est un cas particulier de l'oscillateur de VAN der Pol généralisé avec la méthode des échelles multiples et il obtient le cycle limite à l'ordre 1 qui minimise l'erreur.

L'intérêt de ce travail est de mettre à la disposition des usagers une solution analytique approchée à l'ordre 1 plus proche de la solution exacte de l'équation de VAN der Pol généralisé comme les logiciels MATLAB, MATHEMATICA et autres en numérique.

L'effort à fournir par les usagers serait de ramener l'équation à résoudre sous la forme de l'équation de VAN der Pol ou inversement (en agissant sur les paramètres) afin d'exploiter directement la solution analytique. Pour la validation de la solution une étude comparative est faite. Ainsi nous ramenons l'équation sous la forme de l'équation de l'oscillateur de Rayleigh afin de comparer sa solution à celle trouvée par NAYFEH A. L., (1981) qui était déjà testé.

Dans la première section de ce papier nous avons fait un bref rappel de la théorie de la méthode du groupe de renormalisation, dans la deuxième section nous avons appliqué cette technique à l'oscillateur de VAN der Pol généralisé pour trouver sa solution à l'ordre 1 et enfin dans la dernière partie une étude comparative est faite.

## 2. La méthode du groupe de renormalisation

Dans cette section nous rappelons les grandes lignes de la méthode du groupe renormalisation pour les (EDO). Pour plus détails nous vous renvoyons à (CHEN L. Y. et al., 1994).

On considère une EDO dans ($\mathbb{R}^n$) de la forme

$$\dot{x} = Fx + g(x, t, \epsilon)$$
$$\dot{x} = Fx + \epsilon g_1(x,t) + \epsilon^2 g_2(x,t) + \cdots ; x \in \mathbb{R}^n \qquad (6)$$

où $\epsilon$ est un paramètre positif infiniment petit. Pour ce système, nous supposons que :

1. F soit une matrice carrée $n * n$, diagonalisable à valeurs propres imaginaires ;
2. La fonction $g(x, t, \epsilon)$ soit suffisamment dérivable en t, x et $\epsilon$, le développement en série de puissance de $\epsilon$ soit donné par l'équation (6) ;
3. Chaque $g_i(x, t)$ soit périodique en $t \in \mathbb{R}$ et polynômial en $x$.

Dans un premier temps on applique la méthode de développement simple. On remplace $x$ dans (6) par

$$x(t) = x_0 + \epsilon x_1 + \cdots \qquad (7)$$

Après développement et identification des coefficients de $\epsilon$ on a :

$$\dot{x}_0 = Fx_0 , \qquad (8)$$
$$\dot{x}_1 = F x_1 + G_1(t, x_0), \qquad (9)$$
$$\dot{x}_i = F x_i + G_i(t, x_0 + x_1 + \cdots + x_{i-1}), \qquad (10)$$

où le terme homogène $G_i$ est une fonction régulière de, $x_0, x_{i-1}$. Pour l'instant $G_1, G_2, G_3, G_4$ sont données :

$$G_1(t, x_0) = g_1(x_0, t), \qquad (11)$$
$$G_2(t, x_0, x_1) = \frac{\partial g_1}{\partial x}(x_0, t)x_1 + g_2(x_0, t), \qquad (12)$$
$$G_3(t, x_0, x_1, x_2) = \frac{1}{2}\frac{\partial^2 g_1}{\partial x^2}(x_0, t){x_1}^2 + \frac{\partial g_1}{\partial x}(x_0, t)x_2 + \frac{\partial g_2}{\partial x}(x_0, t)x_1 + g_3(x_0, t). \qquad (13)$$

On a la relation suivante :

$$\frac{\partial G_i}{\partial x_j} = \frac{\partial g_{i-1}}{\partial x_{j-1}} = \frac{\partial g_{i-j}}{\partial x_0} ; i > j \geq 0 \qquad (15)$$

Voir CHIBA H., (2008a) pour la preuve.

Dans les expressions suivantes, on pose $e^{Ft} = X(t)$ et on définit les fonctions $R_1$ et $h_t^i$ sur $\mathbb{R}$ par :

$$R_1(y) = \lim_{t \to \infty} \frac{1}{t} \int_{t_0}^{t} [X(s)^{-1} G_1(s, X(s)y)] ds , \qquad (16)$$

$$h_t^1(y) = X(t) \int_{t_0}^{t} [X(s)^{-1} G_1(s, X(s)y) - R_1(y)] ds, \qquad (17)$$

$$R_i(y) = \lim_{t \to \infty} \frac{1}{t} \int_{t_0}^{t} [X(s)^{-1} G_1\left(s, X(s)y, h_s^1(y), \dots, h_s^{i-1}(y)\right)$$

$$- X(s)^{-1} \sum_{k=1}^{i-1} (Dh_t^k)_y R_{i-k}(y)] ds, i = 2,3 \dots ; \qquad (18)$$

$$h_t^i(y) = X(t) \int_{t_0}^{t} [X(s)^{-1} G_1\left(s, X(s)y, h_s^1(y), \dots, h_s^{i-1}(y)\right)$$

$$- X(s)^{-1} \sum_{k=1}^{i-1} (Dh_t^k)_y R_{i-k}(y) - R_i(y)] ds, \qquad (19)$$

**Proposition** Soit $x_0 = X(t)y$ la solution de l'équation (8) qui a pour condition initiale $y \in \mathbb{R}^n$. Alors pour un temps arbitraire $\zeta \in \mathbb{R}$ et $i = 1, 2, 3 \dots$, la courbe $x_i$ définie par :

$$x_i = x_i(t, \zeta, y) = h_t^i(y) + p_1^i(t, y)(t - \zeta) + p_2^i(t, y)(t - \zeta)^2 + \dots + p_i^i(t, y)(t - \zeta)^i, \quad (20)$$

est la solution de l'équation (10) où les fonctions $p_1^i \dots, p_i^i$ sont données par :

$$p_1^i(t, y) = X(t) R_i(y) + \sum_{k=1}^{i-1} (Dh_t^k)_y R_{i-k}(y), \qquad (21)$$

$$p_j^i(t, y) = \frac{1}{j} \sum_{k=1}^{i-1} \frac{\partial p_{j-1}^k}{\partial y}(t, y) R_{i-k}(y), \quad (j = 2, 3, \dots i - 1); \qquad (22)$$

$$p_j^i(t, y) = \frac{1}{i} \sum_{k=1}^{i-1} \frac{\partial p_{j-1}^k}{\partial y}(t, y) R_{i-k}(y) = \frac{1}{i} \frac{\partial p_{i-1}^{i-1}}{\partial y}(t, y) R_1(y), (23)$$

$$p_j^i(t, y) = 0, (j > i). \qquad (24)$$

Les fonctions $h_t^i$ sont bornées uniformément en t. la solution de l'équation (6) au 1$^{er}$ ordre est donnée par:

$$x(t, \zeta, y) = x_0 + \epsilon x_1 = X(t)y + \epsilon \left(h_t^1(y) + X(t) R_i(y)(t - \zeta)\right) + O(\epsilon^2). \quad (25)$$

Elle est la solution obtenue par développement simple, elle diverge pour les temps longs, d'où la nécessité de sa renormalisation. Elle ne doit pas dépendre de $\zeta$ c'est à dire $(\frac{\partial x}{\partial \zeta} = 0)$, alors

$$0 = X(t) \frac{dy(t)}{dt} + \epsilon \frac{\partial h_t^1}{\partial y} \frac{\partial y(t)}{\partial t} - \epsilon X(t) R_1(y). \qquad (26)$$

On vérifie que (26) admet pour solution :

$$\frac{dy(t)}{dt} = \epsilon R_1(y) + O(\epsilon^2). \tag{27}$$

Soit $y(t)$ une solution de (27), alors la solution de (6) cherchée par la méthode du groupe de renormalisation est donnée par :

$$x(t, t, y) = X(t)y(t) + \epsilon h_t^1(y(t)) + O(\epsilon^2). \tag{28}$$

L'équation (27) est l'équation du groupe de renormalisation de (6). Le calcul pour un ordre supérieur se fait de la même manière et on obtient l'équation du groupe de renormalisation d'ordre m comme suit :

$$\frac{dy}{dt} = \epsilon R_1(y) + \epsilon^2 R_2(y) + \cdots + \epsilon^m R_m(y), \qquad y \in \mathbb{R}^n. \tag{29}$$

En utilisant $h_t^1 \ldots, h_t^m$ définis ci-dessus par (16) et (17), nous définissons la Transformation du groupe de renormalisation d'ordre m (TGR)

$\alpha_t: \mathbb{R}^n \to \mathbb{R}^n, \ y \mapsto \alpha_t(y)$ par

$$\alpha_t(y) = X(t)y + \epsilon h_t^1(y) + \cdots + \epsilon^m h_t^m(y). \tag{30}$$

## 3. Application à l'oscillateur de VAN der Pol généralisé

Dans cette section nous allons appliquer la méthode du groupe de renormalisation à l'oscillateur de VAN der Pol généralisé pour déterminer son équation du groupe de renormalisation et déduire sa solution approchée à l'ordre 1.

L'oscillateur de VAN der Pol généralisé est un système dynamique différentiable à temps continu et à un degré de liberté qui modélise des phénomènes physiques dans les domaines biologique, électronique, médical, musical …

Il est décrit par une coordonnée $x(t)$ vérifiant une équation différentielle faisant intervenir quatre paramètres :

-une pulsation propre $\omega_0 = \sqrt{\frac{k}{m}}$ ;

-deux coefficients positifs $a$ et $b$ appelés paramètres de contrôle ;

-un coefficient de non-linéarité $\epsilon$.

Lorsque $\epsilon = 0$, cet oscillateur se réduit à l'oscillateur harmonique pur.

L'équation différentielle de cet oscillateur s'écrit :

$$\frac{d^2x(t)}{dt^2} + \omega_0^2 x(t) - \epsilon \omega_0 \left[1 - ax^2(t) - b\left(\frac{dx(t)}{dt}\right)^2\right]\frac{dx(t)}{dt} = 0. \tag{31}$$

(31) se réécrit sous la forme :

$$\ddot{x} + x - \epsilon(1 - ax^2 - b\dot{x}^2)\dot{x} = 0 \tag{32}$$

Pour $y = \frac{dx(t)}{dt}, x = (z + \bar{z}) \ et \ y = i(z - \bar{z})$ l'équation (32) devient

$$\begin{cases} \dot{z} = iz + \dfrac{\epsilon}{2}[(z-\overline{z}) - a(z+\overline{z})^2(z-\overline{z}) + b(z-\overline{z})^2] \\ \dot{\overline{z}} = -i\overline{z} - \dfrac{\epsilon}{2}[(z-\overline{z}) - a(z+\overline{z})^2(z-\overline{z}) + b(z-\overline{z})^2] \end{cases} \quad (33)$$

Elle vérifie bien les hypothèses (1-3) avec

$$F = \begin{pmatrix} i & 0 \\ 0 & -i \end{pmatrix}.$$

Les deux équations du système étant identiques, le problème revient à résoudre l'une d'entre elles. Avec

$$z(t) = z_0 + \epsilon z_1 + \cdots \quad (34)$$

On a :

$$\dot{z}_0 = iz_0, \quad (35)$$

$$\dot{z}_1 = i\, z_1 + G_1(t, z_0). \quad (36)$$

A l'ordre zéro on a

$$z_0 = qe^{it} = qZ(t), \quad (37)$$

Avec $q$ la constante d'intégration de (35).

Les équations (16) et (17) donnent respectivement :

$$R_1(q) = \frac{1}{2} q(1 - (a+3b)|q|^2), \quad (38)$$

$$h_t^1(y) = \frac{i}{4}\left[(a-b)\left(q^3 e^{3it} + \frac{1}{2}\overline{q}^3 e^{-3it}\right) + \left((a+3b)q\overline{q}^2 - \overline{q}\right)e^{-it}\right] + C. \quad (39)$$

Avec $i^2 = -1$ et $C$ une constante d'intégration qui sera prise égale à zéro dans la suite pour raison de simplification.

D'après la proposition précédente et les résultats ci-dessus on a :

$$z(t, \zeta, y) = Z(t)q + \epsilon\big(h_t^1(q) + Z(t)R_i(q)(t-\zeta)\big) + O(\epsilon^2), \quad (40)$$

qui diverge pour t long à cause du dernier terme.

En utilisant la notion de renormalisation de la constante d'intégration $\dfrac{\partial z(t,\zeta,q)}{\partial \zeta}\big|_{\zeta=0} = 0$ mentionnée dans la section précédente on a :

$$\begin{cases} z(t,q) = Z(t)q + \epsilon h_t^1(q) + O(\epsilon^2) \\ \dfrac{dq}{d\zeta} = \epsilon R_1(q) + O(\epsilon^2). \end{cases} \quad (41)$$

La première équation de (41) est la solution cherchée par la méthode du groupe de renormalisation et la dernière est l'équation du groupe de renormalisation correspondante.

En posant $q = re^{i\theta(\zeta)}$ avec $x = (z+\overline{z})$ et $y = i(z-\overline{z})$ on trouve :

$$x = 2r\cos(t + \Theta(\zeta)) - \frac{r\epsilon}{2}\sin(t + \Theta(\zeta)) +$$
$$\frac{\epsilon}{2}\left[\left(\frac{b-a}{2}\right)r^3 \sin 3(t + \Theta(\zeta)) + (a + 3b) r^3 \sin(t + \Theta(\zeta))\right] + O(\epsilon^2). \quad (42)$$

L'équation du groupe de renormalisation se transforme en :

$$\begin{cases} \frac{dr}{d\zeta} = \frac{\epsilon r}{2}(1 - (a + 3b)r^2) \\ \frac{d\Theta(\zeta)}{d\zeta} = 0 \end{cases} \quad (43).$$

Il est aisé de prouver que EGR a une orbite périodique stable (le cycle limite) de rayon $r_s = \sqrt{\frac{1}{(a+3b)}}$ avec $(a + 3b) > 0$. Le théorème (Existence of invariant Manifolds) de CHEN L. Y. et al., (1994) nous permet de dire que l'oscillateur de VAN der Pol aussi a une orbite périodique stable dont le rayon $r \neq r_s$ (qui vérifie (43)) évolue et échoue sur le cycle limite $x(r_s, t, \Theta)$ avec $\Theta = cste$ d'après (43).

## 4. Etude comparative

Dans cette section nous allons vérifier la validité de la solution de l'équation de l'oscillateur de VAN der Pol généralisé.

En effet NAYFEH A. H., (1981) a développé plusieurs techniques de détermination de solution approchée des équations différentielles non linéaires.

Dans le chapitre 6 de ce livre intitulé ''Self-excited oscillators '' il a étudié le ''Self-excited Systems'' dont les systèmes sont gouvernés par des équations de la forme :

$$m\frac{d^2 u^*}{dt^{*2}} + ku^* = \mu f^*\left(u^*, \frac{du^*}{dt^*}\right)\frac{du^*}{dt^*}, \quad (44)$$

où $m$ est une masse μ un paramètre positif, $f^*$ est positif pour $u^*$ petit. Pour simplifier le calcul il a utilisé le cas spécial :

$$m\frac{d^2 u^*}{dt^{*2}} + ku^* = \mu\left(1 - \alpha(\frac{du^*}{dt^*})^2\right)\frac{du^*}{dt^*}, \quad (45)$$

Où α est un paramètre positif.

En introduisant les variables sans dimension suivantes :

$$\begin{cases} u = \frac{u^*}{u^*_0} \\ t = t^*\sqrt{\frac{k}{m}} \end{cases} \quad (46),$$

où $u^*_0$ est un déplacement caractéristique et $\omega_0 = \sqrt{k/m}$ est la fréquence propre de

l'oscillateur linéaire, (45) prend la forme :

$$\ddot{u} + u = \epsilon\left(1 - \frac{\alpha u^*_0{}^2 k}{m}\dot{u}^2\right)\dot{u},\qquad(47)$$

où $\epsilon = \frac{\mu}{\sqrt{km}}$ . Pour $u^*_0$ $tel\ que$ $\alpha u^*_0{}^2 k = \frac{m}{3}$ , on obtient la forme standard de (45) dite équation de l'oscillateur de Rayleigh :

$$\ddot{u} + u = \epsilon\left(\dot{u} - \frac{1}{3}\dot{u}^3\right).\qquad(48)$$

NAYFEH A. H. (1981) obtient son cycle limite comme suit :

$$u = 2\cos(t + \beta) + \frac{\epsilon}{12}\sin(3t + 3\beta) + O(\epsilon^2)\ ;\ \boldsymbol{\beta = cste},\qquad(\mathbf{49})$$

Pour $a = 0\ et\ b = \frac{1}{3}$, l'équation de l'oscillateur de VAN der Pol (32) prend la forme :

$$\ddot{x} + x = \epsilon\left(\dot{x} - \frac{1}{3}\dot{x}^3\right).\qquad(50)$$

L'équation de l'oscillateur de VAN der Pol généralisé est ainsi ramenée sous la forme de l'oscillateur de Rayleigh. En remplaçant les valeurs de $a = 0\ et\ b = \frac{1}{3}$, dans (42), on obtient

$$x = 2r\cos(t + \Theta(\zeta)) +$$
$$\frac{\epsilon}{2}\left[-r\sin(t + \Theta) + \frac{r^3}{6}\sin(3t + 3\Theta) + r^3\sin(t + \Theta)\right] + O(\epsilon^2),\qquad(51)$$

avec $r$ solution de (43). En remplaçant $r_s = 1$ dans (51) on retrouve :

$$x = 2r\cos(t + \Theta(\zeta)) + \frac{\epsilon}{12}\sin(3t + 3\Theta) + O(\epsilon^2), \Theta = cste;\qquad(52)$$

(52) est le cycle limite de l'équation de l'oscillateur de VAN der Pol généralisé pour $a = 0\ et\ b = \frac{1}{3}$, qui coïncide avec celui de l'équation de l'oscillateur de Rayleigh. De même pour $a = 0\ et\ b = 3$, on est dans le cas (5) et nous avons retrouvé des résultats conformes à ceux trouvé par CHIBA H., (2008b).

## 5. Conclusion

Nous avons rappelé les grandes lignes de la méthode du groupe de renormalisation pour la résolution des équations différentielles ordinaires (EDO) qui donne en plus de solution, l'équation du groupe de renormalisation (EGR) qui permet la détermination de l'amplitude du cycle limite stable. Une application à, l'oscillateur de VAN der Pol généralisé est faite et la solution approchée trouvée est valable pour tout ordre de la variable $t$. Une étude comparative est faite et elle nous permet de dire que la méthode est efficace et les résultats trouvés sont fiables.

**Remerciements :**

**Références**

ACEVS A., ERCOLAMI N., JONES C., LEGA J. and MOLONEY J., 1994. Introduction to singular perturbation methods nonlinear oscillations. http://math.arizona.edu/~ntna2007/Perturbation Methods.pdf.

BARENBLATT G. I., 1979. Similarity, Self-Similarity, and intermediate Asymptotic (Consultant Bureau, New York) ISBN 10:0306109565/ISBN13.

BENDER C. M. and ORSZAG S. A., 1978. Advanced Mathematical methods for Scientists and Engineers (Mcgraw-Hil, New York).

CART J., 1981. Application of center manifold Theory (Springer, Berlin).

CART J. and MUNCASTER R. M., 1983. *J. Diff. Eq. 50, 260.*

CHEN L. Y., GOLDENFELD N. and OONO Y., 1994. Renormalization group theory for global asymptotic analysis, phys. Rev, lett. 73, no. 10, 1311-15.

CHIBA H., 2008a. $C^1$ approximation de champs de vecteurs basée sur la méthode du groupe de renormalisation, SIAM J. Appl. Dym. syst.

CHIBA H., 2008b. On Renormalization Group Method and its Aplication to Coupled Oscillators. http://www.kurims.kyotou.ac.jp/~kyodo/kokyuroku/contents/ pdf/1616-04.pdf.

CHABI H., Simplified Renormalization Group Method For Ordinary Differential equations. http://www2.math.kyushu-u.ac.jp/~chiba/paper/rg2.pdf

Goldenfeld N., Martin O. and Oono Y., 1989. J. Sci. Comp. 4, 355 .

GOLDENFELD N., MARTIN O., OONO Y. and LIU F., 1990. Phys. Rev. Lett. 64, 1361.

HINCH E. J., 1991. Perturbation Methods (Cambridge University press, Cambridge).

KEVORKIAN and J. D. CODE J. D., 1981. Perturbation Methods in applied Mathematics (Springer-Verlag New York Inc. Edition: 1sted.Softcover of orig.ed).

KUNIHIRO T., EI S., FUJII K., 2000. Renormalisation group method for reduction of evolution equations ; invariant manifolds and envelopes, Ann.Physics 280, n0; 2, 236-298.

LOMOV S. A., 1992. The general theory of Singular Perturbation (American mathematical Society, Providence).

MURDOCK J. A., 1991. Perturbation Theory and Methods (Wiley, New York).

NAYFEH A. H., 1981. *Introduction to perturbation techniques* (Wiley, New York).

PAQUETTE G., CHEN L. Y., GOLDENFELD N., and OONO Y., 1994. phys. Rev.lett. 72, 76.

ROBERTS A. J., 1985. SIAM (Soc. Ind. Appl. Math.) J. MATH. 16, 1243.

SMITH D. R., 1985. Singular Perturbation theory (Cambridge University press, Cambridge).

THOMAS C. B., 2005. The renormalization Group as a method for analyzing differential equation. http://guava.physics.uiuc.edu/~nigel/courses/563/Essays2005/PDF/Butler.pdf.